\documentclass[12pt]{article}
\usepackage{amsmath,amssymb,exscale}  
\input epsf

\def\gappeq{\mathrel{\rlap {\raise.5ex\hbox{$>$}}
{\lower.5ex\hbox{$\sim$}}}}
\def\permil{$\%\raise.20ex\hbox{$_0$}}
\def\lappeq{\mathrel{\rlap{\raise.5ex\hbox{$<$}}
{\lower.5ex\hbox{$\sim$}}}}

\begin{document}
\topmargin -1.0cm
\oddsidemargin -0.8cm
\evensidemargin -0.8cm
\pagestyle{empty}
\begin{flushright}
CERN-TH/2000-081\\
UNIL-IPT-00-06
\end{flushright}
\vspace*{5mm}

\begin{center}

{\Large\bf Bulk fields and supersymmetry in a slice of AdS}\\
\vspace{1.0cm}

{\large Tony Gherghetta$^{a,b}$ and 
Alex Pomarol$^{a,}$\footnote{On leave of absence from
IFAE, Universitat Aut{\`o}noma de Barcelona, 
E-08193 Bellaterra, Barcelona.}}\\
\vspace{.6cm}
{\it {$^{a}$Theory Division, CERN, CH-1211 Geneva 23, Switzerland}}\\
{\it {$^{b}$IPT, University of Lausanne, CH-1015 Lausanne, 
Switzerland}}\\
\vspace{.4cm}
\end{center}

\vspace{1cm}
\begin{abstract}

Five-dimensional models where the bulk is a slice of 
AdS have the virtue of solving the hierarchy problem.
The electroweak scale is generated by a 
``warp'' factor of the induced metric on the brane 
where the standard model fields live.
However, it is not necessary to confine the standard model 
fields on the brane and we analyze the possibility of having 
the fields actually living in the slice of AdS.
Specifically, we study the behaviour of  fermions, 
gauge bosons  and scalars in this geometry and  
their implications on electroweak physics. 
These scenarios  can provide an explanation of the
fermion mass hierarchy by warp factors.
We also consider the case of supersymmetry in the bulk, and 
analyze the conditions on the mass spectrum.
Finally, a model is proposed where the warp factor generates 
a small (TeV) supersymmetry-breaking scale, with the gauge 
interactions mediating the breaking to the scalar sector.

\end{abstract}

\vfill
\begin{flushleft}
CERN-TH/2000-081\\
March 2000
\end{flushleft}
\eject
\pagestyle{empty}
\setcounter{page}{1}
\setcounter{footnote}{0}
\pagestyle{plain}


\section{Introduction}

One the major puzzles in particle physics is the large 
hierarchy of scales that appear in the standard model (SM).
The Planck scale, $M_P\sim 10^{18}$ GeV, 
is much larger than the electroweak scale ($\sim 100$ GeV), and 
even larger than the electron or neutrino mass.
It has recently been realized that the origin
of these hierarchies in scale can be related to the
presence of extra dimensions with nontrivial spacetime 
geometries~\cite{add,rs}.
An explicit example of this is the Randall-Sundrum model \cite{rs}.
This consists of a five-dimensional theory where the extra dimension 
is compactified on an orbifold, and the bulk geometry
is a slice of anti-de Sitter (AdS$_5$) space of length $\pi R$.
Hence, in this theory there are two four-dimensional boundaries.
The induced metric on these boundaries differs by  
an exponential (``warp'') factor, and generates 
two effective scales, $M_P$ and $M_Pe^{-\pi kR}$ 
(where $k$ is the AdS curvature scale of order $M_P$).
Thus, for a moderately large extra dimension, $kR\gappeq 1$, 
one can obtain an exponentially small scale, $M_Pe^{-\pi kR}$,
on one of the boundaries. If the SM fields live on this boundary,
then the electroweak scale can be associated with the effective 
scale on this boundary.  In particular for $kR\sim 12$, this
will be ${\cal O}($TeV$)$ and the brane is referred to as 
the TeV-brane \cite{lr}.

In the model of Ref.~\cite{rs} only gravity propagates in 
the 5D bulk, while the SM fields are on the TeV-brane. 
It is interesting to study other possibilities, 
such as having all the SM fields in the 5D bulk. 
This is the subject of this article, where we will 
study the behaviour of various spin fields
in a slice of AdS$_5$. We will  first analyze the 
non-supersymmetric case. Part of this analysis has been 
carried out previously in the literature.
The role of scalar fields in the bulk was first considered
in Ref.~\cite{gw}. Subsequent studies considered the SM gauge 
bosons~\cite{dhr,gauge}, and fermions~\cite{gn} in the bulk.
The complete SM in the bulk was considered 
in Ref.~\cite{yama,gaba}. We will present here 
a more compact and general analysis, including
for example, the case when scalar fields have boundary mass terms.
As an important new ingredient, we will consider supersymmetry
in the AdS$_5$ slice.
We will derive the conditions necessary for supersymmetry 
and the implications for the particle spectrum.

Armed with the above, we will then 
study the phenomenological consequences
of having the SM fields in the 5D bulk.
We will show that SM fermions and gauge bosons
can propagate in the AdS$_5$ bulk
without conflicting with any experimental data.
However, the Higgs field must arise from a 
Kaluza-Klein excitation that is localized by the AdS metric on 
the TeV-brane~\cite{gw,yama}.
We will show that having the SM fermions in the bulk can explain
the fermion mass hierarchy by means of the metric
warp factor. We will also study
the magnitude of higher-dimensional operators.
These operators are known to be a problem in
theories with a low (TeV) cutoff scale (including in the
original Randall-Sundrum setup~\cite{ddg}),
since they can induce large proton decay rates or
large flavour violating interactions.
These operators can be suppressed
by allowing the fermions to be off the TeV-brane.
This leads to partial success because flavour violation can be 
suppressed to the desirable level, but this is 
not the case for proton decay.

The inclusion of supersymmetry allows for new alternatives.
The Higgs field can now be delocalized from the TeV-brane,
together with the fermions.
This delocalization is possible thanks to supersymmetry that 
protects the Higgs massless mode far from the TeV-brane.
If fermions are also distant from the TeV-brane, then 
proton decay can be sufficiently suppressed, 
while at the same time allowing for the fermion masses to 
be generated. In fact, this model  at low energy
resembles the ordinary MSSM.
Nevertheless, it does have important differences.
For instance, the boundary TeV-brane can now be 
the source of supersymmetry breaking and a window 
to ``Planckian'' physics. 
Gauge superfields living in the bulk will be responsible
for mediating TeV-brane physics to fields that are far away from it.
For example, the gauge interactions can be the 
messengers of supersymmetry breaking to the quark 
and lepton sector, guaranteeing a universal scalar mass spectrum.
Similar scenarios were considered in the case of 
a flat extra dimension. This was done either by breaking 
supersymmetry in the bulk~\cite{bulk} or on a distant 
brane~\cite{brane}. In Refs.~\cite{bulk}, a large (TeV) extra dimension
was required that implied a cutoff (string) scale of $\cal O$(TeV).
In Refs.~\cite{brane}, a small (Planckian) extra dimension was 
advocated, but in this case the resultant phenomenology 
was similar to the gravity-mediated models of supersymmetry-breaking
proposed earlier with no new low-energy implications.
Nevertheless, the possibility here of mediating supersymmetry-breaking
by five-dimensional gauge bosons leads to a very different scenario
with respect to the previous examples.
As we will see, we can combine the good implications 
of having a large ($\sim M_P$) cutoff scale 
(such as small proton decay, small  neutrino masses), 
with the new phenomenology of having an extra dimension
(Kaluza-Klein states at collider physics, quantum gravity effects 
at the TeV).

It is important to point out why the slice of AdS$_5$ 
is interesting and different
from other compactifications \cite{otherc}.
Theories that solve the hierarchy problem with 
other compactifications \cite{add,otherc}
usually require a large compactification radius 
(or, equivalently, a large volume in the extra dimensions).
Therefore, if SM fields  propagate in these extra dimensions,
the couplings become too weak.

Finally, three important remarks are in order. 
Even though we will be considering 
fields other than the graviton living in the 
5D bulk, we will assume that the AdS metric 
is not modified by the presence of these bulk fields.
In other words, the back-reaction from the bulk
fields is neglected.
Second, there have been recent unsuccessful
attempts at deriving the scenario of Ref.~\cite{rs}
from a five-dimensional supergravity Lagrangian~\cite{ads}.
While there is no general proof that a description from
a more fundamental theory does not exist, we will not have 
anything more to add here.
Further studies must still be carried out to settle this issue.
We will assume that the solution of Ref.~\cite{rs} exists
and study its compatibility  with supersymmetry. Finally, to 
obtain the AdS slice, the TeV-brane in Ref.~\cite{rs}
requires a negative tension. This will violate the 
weak-energy condition~\cite{witt}. Nevertheless,
it is conceivable that a setup, like we are considering, 
is possible without violating the weak-energy condition~\cite{witt}, 
especially given the
fact that the fundamental theory is most likely to
be supersymmetric.

The article is organized in the following way.
In Section~2, we introduce the compactification scenario
of Ref.~\cite{rs}, and derive the Kaluza-Klein 
decomposition of fields with different spins.
In Section~3, we study the conditions necessary 
for supersymmetry in a slice of AdS$_5$. In particular, we
derive the mass spectrum and wavefunctions of the 
gauge supermultiplet and hypermultiplet.
The possibility of having the SM fermions
and gauge bosons in the 5D bulk is studied in Section~4, where
particular attention is paid to constraints from electroweak data.
This will lead us to consider supersymmetric models, together with 
their resulting phenomenological predictions.
Our conclusions and final comments appear in Section~5.

\section{Bulk fields in a slice of AdS$_5$}

We will consider the scenario of Ref.~\cite{rs}, 
based on a non-factorizable geometry with one extra dimension.
In this scenario, the fifth dimension $y$ is compactified on an 
orbifold, $S^1/\mathbb{Z}_2$ of radius $R$, with 
$-\pi R\leq y\leq\pi R$. The orbifold fixed points at, 
$y^\ast=0$ and $y^\ast=\pi R$ are also the locations of two 3-branes, 
which form the boundary of the five-dimensional space-time.
Consequently, the classical action for this configuration is given by
\begin{equation}
    S=-\int d^4 x \int_{-\pi R}^{\pi R} dy \sqrt{-g} 
       \left[ \Lambda +\frac{1}{2} M_5^3 {\cal R}
        +\delta(y) \left( \Lambda_{(0)} + {\cal L}_{(0)} \right)
        +\delta(y-\pi R) \left( \Lambda_{(\pi R)} + {\cal L}_{(\pi R)} 
       \right) \right]~,
\end{equation}
where $M_5$ is the five-dimensional
Planck mass scale and ${\cal R}$ is the five-dimensional 
Ricci scalar, 
constructed from the five-dimensional metric $g_{MN}$, with 
the 5D coordinates labelled by capital Latin letters, $M=(\mu,5)$. 
The cosmological constants in the bulk and boundary are $\Lambda$, 
and $\Lambda_{(y^\ast)}$, respectively. 

A solution to the five-dimensional Einstein's equations, which respects 
four-dimensional Poincare invariance in the $x^\mu$ directions, 
is given by~\cite{rs}
\begin{equation}
\label{metric}
        ds^2=e^{-2\sigma}\eta_{\mu\nu}dx^\mu dx^\nu+dy^2\, ,
\end{equation}
where 
\begin{equation}
        \sigma=k|y|~,
\end{equation}
and $1/k$ is the AdS curvature radius. The four-dimensional metric is 
$\eta_{\mu\nu}={\rm diag}(-1,1,1,1)$ with $\mu=1,..,4$. 
The metric solution (\ref{metric}) 
is valid provided that the bulk and boundary cosmological constants are 
related by
\begin{eqnarray}
\label{cosmorel}
     \Lambda&=&-6M_5^3k^2~,\nonumber\\
     \Lambda_{(0)}&=&-\Lambda_{(\pi R)}=-\frac{\Lambda}{k}~.
\end{eqnarray}
The four-dimensional reduced Planck scale $M_P$
is related to $M_5$ in the following way:
\begin{eqnarray}
M_P^2=\frac{M_5^3}{k}\left(1-e^{-2\pi kR}\right)\, .
\end{eqnarray}
From the form of the metric solution (\ref{metric}), 
the space-time between the two branes 
located at $y^\ast=0$ and $y^\ast=\pi R$ is simply 
a slice of AdS$_5$ geometry. 
The effective mass scale on the brane at $y^\ast=0$ is the
Planck scale $M_P$, and we will refer to this 3-brane as the Planck-brane.
Similarly, at $y^\ast=\pi R$, the effective mass scale is $M_P e^{-\pi kR}$, 
which 
will be associated with the TeV scale provided $kR\simeq 12$. This 3-brane will
be referred to as the TeV-brane. It is also interesting to note that the  
relations (\ref{cosmorel}) arise in the effective 
five-dimensional Horava-Witten theory~\cite{losw} when 
the Calabi-Yau moduli are fixed.

So far we have only assumed gravity 
to be present in the five-dimensional bulk. 
We will now study a U(1) gauge field, $V_M$,  
a complex scalar, $\phi$, and a Dirac fermion, $\Psi$,
living in a slice of AdS$_5$ given by the metric Eq.~(\ref{metric}).
The five-dimensional (5D) bulk action, $S_5$, has kinetic energy and 
mass terms given by
\begin{equation}
\label{kin}
   S_5=-\int d^4x\int dy\sqrt{-g}\, \Bigg[\frac{1}{4g^2_5}F^2_{MN}+
     \left|\partial_M\phi\right|^2+i\bar{\Psi}\gamma^MD_M\Psi
     +m^2_\phi|\phi|^2+im_\Psi\bar{\Psi}\Psi\Bigg]\, ,      
\end{equation}
where $g={\rm det}(g_{MN})$.
The U(1) gauge field strength is 
$F_{MN}=\partial_M V_N-\partial_N V_M$ and in curved space the covariant 
derivative is $D_M=\partial_M+\Gamma_M$, where $\Gamma_M$ is the
spin connection. In particular, for the metric defined by Eq.~(\ref{metric}),
we have
\begin{equation}
\Gamma_\mu= \frac{1}{2}\gamma_5\gamma_\mu\frac{d\sigma}{dy} 
\ \ \ \ {\rm and}\ \ \ \ \Gamma_5=0\, .
\end{equation}
The gamma matrices, $\gamma_M=(\gamma_\mu,\gamma_5)$ are defined in curved 
space as $\gamma_M=e^\alpha_M\gamma_\alpha$,
where $e^\alpha_M$ is the vierbein and $\gamma_\alpha$ are the Dirac matrices
in flat space.

Under the $\mathbb{Z}_2$ parity the boson fields 
can be defined either odd or even (depending on their interactions).
For the fermion $\Psi$, the $\mathbb{Z}_2$ transformation is given 
by $\Psi(-y)=\pm\gamma_5\Psi(y)$, where the arbitrariness of the sign 
can only be determined by the fermion interactions. We then have that 
$\bar{\Psi}\Psi$ is odd under the $\mathbb{Z}_2$ symmetry, and consequently 
the Dirac mass parameter $m_\Psi$ must also be odd. 
This nontrivial transformation of the fermion mass parameter
implies that $m_\Psi$ must arise from an underlying scalar 
field that receives a vacuum expectation value (VEV) with an 
odd `kink' profile. In other words, the 
vacuum configuration of the scalar field can be thought of as 
an infinitely thin 
domain wall. This is very similar to the background 3-form field
in the dimensional reduction of Horava-Witten theory~\cite{losw}.
Furthermore, note that a Dirac mass term cannot 
be induced on the boundaries, since $\bar{\Psi}\Psi$ vanishes there. 
On the other hand, a Majorana mass term for the fermion
can be added, either in the bulk or on the boundaries. This term will
necessarily break supersymmetry, and we will comment on this
in section~\ref{pheno}.
The scalar mass term 
is even under the $\mathbb{Z}_2$ symmetry and can be either
a bulk or boundary term. Therefore, the mass parameters of the scalar
and fermion fields can be parametrized as 
\footnote{We are assuming that 
the magnitude of the boundary mass for the 
scalar is the same on the two boundaries, but with opposite sign. 
As we will see, this is required
by supersymmetry.
The generalization to different 
masses for different boundaries can be easily obtained from the analysis here.}
\begin{eqnarray}
\label{masses}
       m^2_\phi&=&ak^2+b\sigma^{\prime\prime}\, ,\nonumber\\
       m_\Psi&=&c\sigma^\prime\, ,
\end{eqnarray}
where $a,b$ and $c$ are arbitrary dimensionless parameters, 
and the derivatives are defined
as
\begin{eqnarray}
\label{derivatives}
        \sigma^\prime&=&\frac{d\sigma}{dy}=k\epsilon(y)~,\nonumber\\  
        \sigma^{\prime\prime}&=&\frac{d^2\sigma}{dy^2}=2k
        \left[\delta(y)-\delta(y-\pi R)\right]~.
\end{eqnarray} 
The step function $\epsilon(y)$ is defined as being $1$ ($-1$)
for positive (negative) $y$, and $\delta(y)$ is the Dirac delta-function.

The equations of motion for the 
gauge, scalar and fermion fields are respectively given by
\begin{eqnarray}
\label{eq.motion}
      \partial_M(\sqrt{-g}g^{MN}g^{RS}F_{NS})&=&0\, ,\nonumber\\
      \frac{1}{\sqrt{-g}}\partial_M(\sqrt{-g}g^{MN}\partial_{N}\phi)
      -m^2_\phi\phi&=&0~, \nonumber\\
      (g^{MN}\gamma_M D_N+m_\Psi)\Psi&=&0~.
\end{eqnarray}
Using the metric of Eq.~(\ref{metric}), one can write a general
second-order differential equation \footnote{For the gauge field we 
work, as in the graviton case \cite{rs},
 in the gauge where $V_5=0$ together with   the constraint
$\partial_\mu V^\mu =0$.}
\begin{equation}
\label{difeq}
      \left[e^{2\sigma}\eta^{\mu\nu}\partial_\mu\partial_\nu+
       e^{s\sigma}\partial_5(e^{-s\sigma}\partial_5)-M^2_\Phi\right]
       \Phi(x^\mu,y)=0\, ,
\end{equation}
where $\Phi=\{V_\mu,\phi,e^{-2\sigma}\Psi_{L,R}\}$, $s=\{2,4,1\}$ and 
$M^2_\Phi=\{0,ak^2+b\sigma^{\prime\prime}, 
c(c \pm 1)k^2 \mp c\sigma^{\prime\prime}\}$. 
For the fermion, we have introduced the exponential factor, $e^{-2\sigma}$,
which takes into account the spin connection, and we have explicitly 
separated the left and right components 
(defined by $\Psi_{L,R}=\pm\gamma_5\Psi_{L,R}$).

\subsection{Kaluza-Klein decomposition}

Let us decompose the 5D fields as
\begin{equation}
\label{Kaluza-Klein}
        \Phi(x^\mu,y)={1\over\sqrt{2\pi R}}\sum_{n=0}^\infty 
\Phi^{(n)}(x^\mu)f_n(y)\, ,  
\end{equation}
where the Kaluza-Klein modes $f_n(y)$ obey the orthonormal condition
\begin{equation}
\label{orthonorm}
      {1\over 2\pi R}\int^{\pi R}_{-\pi R}dy\, e^{(2-s)\sigma}f_n(y) 
       f_m(y)=\delta_{nm}~.  
\end{equation}
Thus from Eq.~(\ref{difeq}), the Kaluza-Klein 
eigenmodes $f_n(y)$ satisfy the differential equation
\begin{equation}
\label{kkequation}
      \left[-e^{s\sigma}\partial_5 (e^{-s\sigma}\partial_5)
       +\widehat M^2_\Phi\right]f_n=e^{2\sigma}m^2_n f_n\, .
\end{equation}
In Eq.~(\ref{kkequation}), we have defined the mass-squared parameter,
$\widehat M^2_\Phi
=\{0,ak^2,c(c\pm 1)k^2\}$,
without the boundary
mass terms proportional to $\sigma^{\prime\prime}$. These terms will be
considered later when we impose the boundary conditions on the solutions.
The corresponding eigenfunctions $f_n$, obtained by solving
Eq.~(\ref{kkequation}), are
\begin{equation}
\label{solution}
     f_n(y)=\frac{e^{s\sigma/2}}{N_n}\left[J_\alpha(\frac{m_n}{k}e^{\sigma})
     +b_{\alpha}(m_n)\, Y_\alpha(\frac{m_n}{k}e^{\sigma})\right]\, ,
\end{equation}
where $m_n$ is the mass of the Kaluza-Klein excitation $\Phi^{(n)}$ and
the normalization factor $N_n$ and coefficients $b_\alpha(m_n)$ are constants. 
The Bessel functions $J_\alpha$ and $Y_\alpha$ are of order
$\alpha=\sqrt{(s/2)^2+\widehat M^2_\Phi/k^2}$. Using the orthonormal 
condition Eq.~(\ref{orthonorm}), an expression for
the normalization constant $N_n$ is
\begin{equation}
\label{normn}
      N_n^2 = {1\over\pi R} \int_0^{\pi R} dy\, e^{2\sigma} 
      \left[ J_\alpha(\frac{m_n}{k}e^{\sigma}) + b_\alpha(m_n) 
      Y_\alpha(\frac{m_n}{k}e^\sigma)\right]^2~.  
\end{equation}
In the limit $m_n \ll k $ and $kR\gg 1$ an approximate expression
for the normalization constant is
\begin{equation}
     N_n \simeq {e^{\pi kR}\over \sqrt{2\pi kR}} J_\alpha ({m_n\over k}
     e^{\pi kR})\simeq {e^{\pi kR/2}\over \sqrt{\pi^2 R m_n}}~.
\end{equation}
Note that in this limit the normalization constant, $N_n$,  
does not depend on the particular type of field i.e. the parameter $s$.
The Kaluza-Klein masses $m_n$ and the coefficients $b_{\alpha}(m_n)$ 
are determined
by imposing the boundary conditions on the solution (\ref{solution}).
The appropriate boundary condition depends on whether the field is odd or 
even under the orbifold $\mathbb{Z}_2$ symmetry.

\noindent {\bf Even fields:}

If the 5D field is even under the $\mathbb{Z}_2$ symmetry, then we have
\begin{equation}
\label{eq:even}
     f_n(y)\rightarrow f_n(|y|)\, .
\end{equation}
The derivative of $f_n$ must be either discontinuous (proportional 
to $\epsilon(y)$) or continuous on the boundaries depending on 
whether or not the mass-squared parameter, $M_\Phi^2$ in Eq.~(\ref{difeq}),
has delta function terms, $\sigma^{\prime\prime}$.
Thus we must impose
\begin{equation}
\label{boun:even}
 \left.\left(\frac{df_n}{dy} - r \sigma^\prime f_n\right)
\right|_{0,\pi R}= 0\, 
,
\end{equation}
where the parameter $r$ has the values $r=\{0,b,\mp c\}$ for 
$\Phi=\{V_\mu,\phi,e^{-2\sigma}\Psi_{L,R}\}$, respectively.
Imposing Eq.~(\ref{boun:even}) gives rise to the two equations
\begin{eqnarray}
\label{beven}
    b_{\alpha}(m_n)&=&-\frac{(-r+s/2)J_\alpha(\frac{m_n}{k})+\frac{m_n}{k}
      J^\prime_\alpha(\frac{m_n}{k})}{(-r+s/2)Y_\alpha(\frac{m_n}{k})
      +\frac{m_n}{k}Y^\prime_\alpha(\frac{m_n}{k})}\, ,\\
    b_{\alpha}(m_n)&=&b_{\alpha}(m_ne^{\pi kR})\, .
\end{eqnarray}
These two conditions determine the values of $b_\alpha$ and $m_n$. 
In the limit that $m_n \ll k$ 
and $kR \gg 1$ the Kaluza-Klein mass solutions for $n=1,2,\dots$
and $\alpha> 0$ are
\begin{equation}
\label{evenmn}
   m_n\simeq (n+\frac{\alpha}{2}-\frac{3}{4})\pi k e^{-\pi kR}~.
\end{equation}
The approximate mass formula (\ref{evenmn}) becomes more exact, 
the larger the value of $n$.

\newpage
\noindent {\bf Odd fields:}

If the 5D field is odd under the $\mathbb{Z}_2$ symmetry, then we have
\begin{equation}
\label{eq:odd}
       f_n(y)\rightarrow \frac{\sigma^\prime}{k}f_n(|y|)\, .
\end{equation}
The continuity of $f_n$ at the boundaries implies that  
\begin{equation}
\label{boun:odd}
      f_n\Big|_{0,\pi R}=0\, ,  
\end{equation}
and consequently
\begin{eqnarray}
\label{bodd}
      b_{\alpha}(m_n)&=&-\frac{J_\alpha(\frac{m_n}{k})}
        {Y_\alpha(\frac{m_n}{k})}\, ,\\
      b_{\alpha}(m_n)&=&b_{\alpha}(m_ne^{\pi k R})\, .
\end{eqnarray}
In this case one can check that the derivative of $f_n$ is continuous on 
the boundaries and does not lead to further conditions. As in the even case, an
approximate solution for the Kaluza-Klein tower in the limit
that $m_n \ll k$ and $kR \gg 1$ is
\begin{equation}
\label{oddmn}
   m_n\simeq (n+\frac{\alpha}{2}-\frac{1}{4})\pi k e^{-\pi kR}~,
\end{equation}
where $n=1,2,\dots$ and $\alpha > 0$. Note that the difference 
in the approximate mass formulae between the even and odd mass 
solutions is $\pi/2$.

In the case of fermion fields one should note that the even and odd functions
are related by coupled first-order differential equations
\begin{eqnarray}
     \gamma^\mu\partial_\mu\hat\Psi_R +\partial_5\hat\Psi_L
     +m_\Psi \hat\Psi_L &=&0~, \\
     \gamma^\mu\partial_\mu\hat\Psi_L -\partial_5
     \hat\Psi_R +m_\Psi \hat\Psi_R &=&0~,
\end{eqnarray}
where $\hat\Psi = e^{-2\sigma} \Psi$. Thus,
the even and odd fermion fields must also satisfy these equations.

\section{Supersymmetry in a slice of AdS$_5$}
\label{sectionsusy}

An AdS space is compatible with supersymmetry~\cite{susyads}.
However, in contrast to the flat space case, AdS supersymmetry requires that
different fields belonging to the same supersymmetric multiplet 
have different masses. This is because the momentum operator, $P$,
in AdS space does not commute with the supersymmetric charges
and $P^2$ is not a Casimir operator.

The supermultiplet mass-spectrum has been derived in Ref.~\cite{shuster} for a 
five-dimensional AdS space. Here we will extend the analysis to the case 
where the fifth dimension is compactified on an orbifold $S^1/\mathbb{Z}_2$. 
In the next section,
we will see that the supermultiplet mass-spectrum in a slice of AdS$_5$
will have interesting phenomenological implications.

\subsection{Supergravity multiplet}

The on-shell supergravity multiplet consists of the vierbein $e^\alpha_M$,
the graviphoton $B_M$ and two symplectic-Majorana~\footnote{We will 
follow the conventions of Ref.~\cite{peskin}  (except for our metric
convention $\eta_{\mu\nu}=(-1,1,1,1)$) where the
two symplectic-Majorana spinors, $\psi^i$ and $\chi^i$, satisfy
$\bar\psi^i\gamma^M\dots\gamma^P\chi^j=-\epsilon^{ik}\epsilon^{jl}
\bar\chi^l\gamma^P\dots\gamma^M\psi^k$.}
gravitinos $\Psi^i_M$ $(i=1,2)$.
The index $i$ labels the fundamental representation of the SU(2) automorphism 
group of the $N=1$ supersymmetry algebra in five dimensions. 
In a slice of AdS$_5$, the supergravity  Lagrangian has extra 
terms proportional to the cosmological constants:
\begin{equation}
\label{gravity}
    S_5=-\frac{1}{2}\int d^4x\int dy\sqrt{-g}\, \Bigg[M_5^3
    \Big\{{\cal R}+i\bar\Psi^i_M\gamma^{MNR}D_N\Psi_{R}^i
    -i\frac{3}{2}\sigma^\prime \bar\Psi^i_M\sigma^{MN}(\sigma_3)^{ij}
    \Psi_{N}^j\Big\}+2\Lambda-\frac{\Lambda}{k^2}\sigma^{\prime\prime}
    \Bigg]\, ,
\end{equation}
where $\gamma^{MNR}\equiv \sum_{\rm perm} (-1)^p\gamma^M\gamma^N\gamma^R/3!$
and $\sigma^{MN}=[\gamma^M,\gamma^N]/2$.
In Eq.~(\ref{gravity}) we do not show the dependence on $B_M$,
since in the AdS$_5$  background we set $B_M=0$.
In order to respect supersymmetry in AdS$_5$, the supersymmetric 
transformation of the gravitino must be changed, with respect
to the supergravity case with no cosmological constant, 
in the following way
\begin{equation}
\label{sugratrans}
   \delta \Psi^i_M=D_M\eta^i+\frac{\sigma^\prime}{2}\gamma_M(\sigma_3)^{ij}
    \eta^j\, ,
\end{equation}
where  $\sigma_3={\rm diag}(1,-1)$ and 
the symplectic-Majorana spinor $\eta^i$ is the supersymmetric parameter.
Without loss of generality, we have defined the 
$\mathbb{Z}_2$ transformation of the symplectic-Majorana spinor as
\begin{equation}
        \eta^i(-y)=(\sigma_3)^{ij} \gamma_5 \eta^j(y)\, .
\end{equation}

The condition that the AdS$_5$ background does not break supersymmetry
is $\delta\Psi_M^i=0$, and using Eq.~(\ref{sugratrans}) this leads to the
Killing spinor equation
\begin{equation}
\label{killing}
  D_M\eta^i=-\frac{\sigma^\prime}{2}\gamma_M(\sigma_3)^{ij}\eta^j\, .
\end{equation}
In a non-compact five-dimensional AdS space this condition is always 
fulfilled. However in the orbifold compactification, the boundary terms require
an extra condition to be satisfied, namely
\begin{equation}
  \label{extrakill}
  \gamma_5\eta^i=(\sigma_3)^{ij}\eta^j~.
\end{equation}
This condition implies that only half of the 5D supersymmetric charges 
are preserved. Therefore after compactification, one has in 4D a $N=1$
supersymmetric theory instead of $N=2$.

\subsection{Vector supermultiplet}

The on-shell field content of the vector supermultiplet
is $\mathbb{V}=(V_M,\lambda^i,\Sigma)$ 
where  $V_M$ is the gauge field, $\lambda^i$ is a symplectic-Majorana
spinor, and $\Sigma$ is a real scalar in the adjoint representation.
For simplicity we will consider a U(1) gauge group. The action
has the form
\begin{equation}
\label{kinvector}
       S_5=-\frac{1}{2}
\int d^4x\int dy\sqrt{-g}\, \Bigg[\frac{1}{2g^2_5}F^2_{MN}+
\left(\partial_M\Sigma\right)^2+i\bar{\lambda^i}\gamma^MD_M\lambda^i     
+m^2_\Sigma\Sigma^2+im_\lambda\bar{\lambda^i}(\sigma_3)^{ij}\lambda^j
\Bigg]\, .
\end{equation}
In flat space, supersymmetry requires that $m_\Sigma = m_\lambda=0$. 
This is to be contrasted with an AdS$_5$ background, where fields in the 
same supermultiplet must have different masses~\cite{shuster}. By requiring
Eq.~(\ref{kinvector}) to be invariant under the 
supersymmetric transformations
\begin{eqnarray}
\delta V_M&=&-i\bar\eta^i\gamma_M\lambda^i~,\nonumber\\
\delta\Sigma&=&\bar\eta^i\lambda^i~,\nonumber\\
\delta\lambda^i&=&(-\sigma^{MN}F_{MN}+i\gamma^M\partial_M\Sigma)\eta^i
-2i\sigma^\prime\Sigma(\sigma_3)^{ij}
\eta^j\, ,
\end{eqnarray}
where $\eta^i$ satisfies the Killing equation (\ref{killing}) and the
condition (\ref{extrakill}),
one finds that in a slice of AdS$_5$, the five-dimensional 
masses of the scalar and spinor fields in the vector supermultiplet 
must be
\begin{eqnarray}
\label{v:susycon}
       m^2_\Sigma&=&-4k^2+2\sigma^{\prime\prime}\, ,\nonumber\\
       m_\lambda&=&\frac{1}{2}\sigma^\prime\, .
\end{eqnarray}
The Kaluza-Klein decomposition can be obtained easily from the analysis of 
the previous sections. From Eq.~(\ref{v:susycon}) and  Eq.~(\ref{masses}) 
we have
\begin{equation}
\label{v:abc}
            a=-4\ , \ \  b=2\ , \ \  {\rm and}\ \ c=\frac{1}{2}\, .
\end{equation}
Using Eq.~(\ref{v:abc}), we find that $\alpha=1$ for 
$V_\mu$ and $\lambda^1_L$, while $\alpha=0$ for $\Sigma$ and $\lambda^2_L$.
If we assume that $V_\mu$ and $\lambda^1_L$ are even, while $\Sigma$ and 
$\lambda^2_L$ are odd, then the Kaluza-Klein masses
are determined by the equation
\begin{equation}
      \frac{J_0(\frac{m_n}{k})}{Y_0(\frac{m_n}{k})}=
      \frac{J_0(\frac{m_n}{k}e^{\pi kR})}{Y_0(\frac{m_n}{k}e^{\pi kR})}\, .
\end{equation}
Thus, even though the values of $\alpha$ are different, we find as expected
that all fields of the supermultiplet have identical Kaluza-Klein masses.
In fact, the approximate solution for the mass of the Kaluza-Klein modes 
with $n=1,2,\dots$ is given by~\cite{gauge}
\begin{equation}
    m_n\simeq (n-\frac{1}{4})\pi k e^{-\pi kR}~,
\end{equation}
and is consistent with the mass formulae (\ref{evenmn}) and (\ref{oddmn}).
The even fields $V_\mu$ and $\lambda^1_L$ will 
have a massless mode with the following $y$ dependence:
\begin{eqnarray}
\label{eq:massless}
       V_\mu(x,y)&=&\frac{1}{\sqrt{2\pi R}} V^{(0)}_\mu(x)+\dots\, ,\nonumber\\
       \lambda^1_L(x,y)&=&\frac{e^{3\sigma/2}}{\sqrt{2 \pi R}}
      \lambda^{1\, (0)}_L(x)+\dots\, .
\end{eqnarray}
Notice that the zero modes have the proper $y$ dependence to make the 
4D kinetic terms of these modes in the action
\begin{equation}
\label{kinaction}
    -\int d^4x \frac{1}{2\pi R}\int_{-\pi R}^{\pi R} dy \sqrt{-g} 
    \left[\frac{1}{2g^2_5}
\partial^\nu V^{\mu\, (0)}(x)\partial_\nu V^{(0)}_\mu(x)
    +ie^{3\sigma}\bar\lambda^{1\, (0)}_L(x)
    \gamma^\mu\partial_\mu \lambda^{1\, (0)}_L(x)\right]~,
\end{equation}
invariant under the conformal transformation $g_{\mu\nu}\rightarrow 
e^{2\sigma}g_{\mu\nu}$. This implies that the action (\ref{kinaction}) in the 
background metric (\ref{metric}) is independent of the $y$ coordinate. 
Consequently, 
the zero modes are not localized by the AdS space-time and they 
behave like zero modes in flat space. 
Furthermore, their couplings to the two boundaries are of equal strength.

The odd fields $\Sigma$ and $\lambda^2_L$ do not have massless modes because 
this is not consistent with the orbifold condition. Therefore, the massless 
sector from $V_\mu$ and $\lambda^1_L$ forms an $N=1$ supersymmetric vector 
multiplet.

\subsection{Hypermultiplet}

The hypermultiplet consists of $\mathbb{H}=(H^i,\Psi)$  
where $H^i$  are  two complex scalars and $\Psi$ is a Dirac fermion.
The action has the form
\begin{equation}
\label{kinhyper}
   S_5=-\int d^4x\int dy\sqrt{-g}\, \Bigg[
     \left|\partial_M H^i \right|^2+i\bar{\Psi}\gamma^MD_M\Psi
     +m^2_{H^i}|H^i|^2+im_\Psi\bar{\Psi}\Psi\Bigg]\, .   
\end{equation}
Invariance under the supersymmetric transformations
\begin{eqnarray}
      \delta H^i&=&\sqrt{2}i\epsilon^{ij}\bar\eta^j\Psi\, ,\nonumber\\
      \delta\Psi&=&\sqrt{2}\left[\gamma^M\partial_M H^i\epsilon^{ij}-
      \frac{3}{2}\sigma^\prime H^i (\epsilon\sigma_3)^{ij}
      -m_\Psi H^i\epsilon^{ij}\right]\eta^j\, ,
\end{eqnarray}
where $\eta^i$ satisfies the Killing equation (\ref{killing}) and 
the condition (\ref{extrakill}),
requires that the five-dimensional masses of the scalars and fermion satisfy
\begin{eqnarray}
\label{h:susycon}
       m^2_{H^{1,2}}&=&(c^2\pm c-\frac{15}{4})k^2
       +\left(\frac{3}{2}\mp c\right)\sigma^{\prime\prime}\, ,\nonumber\\
       m_\Psi&=&c\sigma^\prime\, ,
\end{eqnarray}
where $c$ remains an arbitrary dimensionless parameter. Using 
Eq.~(\ref{masses}) we can identify
\begin{equation}
\label{h:abc}
     a=c^2\pm c-\frac{15}{4} \quad {\rm and} \quad  b=\frac{3}{2}\mp c\, .
\end{equation}
Thus, we find that $\alpha=| c+1/2|$ for 
$H^1$ and $\Psi_L$, and $\alpha=| c-1/2|$ for $H^2$ and $\Psi_R$.
Assuming that $H^1$ and $\Psi_L$ are even, while $H^2$ and $\Psi_R$ are odd,
then the Kaluza-Klein masses (identical for both the even and 
odd modes) are determined by the equation
\begin{equation}
       \frac{J_{| c+1/2|}(\frac{m_n}{k})}{Y_{| c+1/2|}(\frac{m_n}{k})}
      =\frac{J_{| c+1/2|}(\frac{m_n}{k}e^{\pi kR})}{Y_{| c+1/2|}
      (\frac{m_n}{k}e^{\pi kR})}\, .
\end{equation}
In fact using Eq.(\ref{evenmn}) for the even modes, and Eq.(\ref{oddmn}) 
for the odd modes, we can again see that the odd and even Kaluza-Klein modes,
$n=1,2,\dots$ have identical masses which 
are approximately given by
\begin{equation}
     m_n\simeq (n+\frac{ c}{2}-\frac{1}{2}) \pi k e^{-k\pi R}~.
\end{equation}
The massless modes of $H^1$ and $\Psi_L$ are given by
\begin{eqnarray}
\label{eq:masslessH}
      H^1(x,y)&=&\frac{e^{(3/2- c)\sigma}}{\sqrt{2\pi R}N_0}
      H^{1\, (0)}(x)+\dots\, ,\nonumber\\
      \Psi_L(x,y)&=&\frac{e^{(2- c)\sigma}}
      {\sqrt{2\pi R}N_0}\Psi^{(0)}_L(x)+\dots\, ,
\end{eqnarray}
where 
\begin{equation}
     N^2_0 = \frac{e^{2\pi kR(1/2- c)}-1}{2\pi kR(1/2- c)}~.
\end{equation}
When $ c=1/2$, we have $N_0=1$ which signals the conformal 
limit, where
the kinetic terms are again independent of the $y$ coordinate. 
Thus, as explained in the vector multiplet case, the massless modes
are not localized by the AdS space and will couple to the two boundaries 
with equal strength.
For values of $ c<1/2$, the AdS space will localize the massless modes 
towards the boundary $y^\ast=\pi R$. 
As $ c$ becomes more negative, the localization at the $y^\ast=\pi R$ 
boundary becomes increasingly more effective.
On the other hand when $ c>1/2$, the massless modes will be localized 
towards the $y^\ast=0$ boundary. In this case the 
localization becomes more effective the larger the value of $ c$.

The odd fields do not have a massless mode, since this is not consistent 
with the orbifold condition.
Therefore, as in the case of the vector multiplet, the massless sector 
of the hypermultiplet, (from $H^1$ and $\Psi_L$), forms an 
$N=1$ supersymmetric chiral multiplet.

\section{Phenomenological implications}
\label{pheno}

In the original proposal in Ref.~\cite{rs}, the SM fields were 
assumed to be confined on the TeV-brane. 
The question naturally arises then of 
whether the SM fields can actually live in the bulk. We will
first consider models with bulk fields but without supersymmetry. A
non-supersymmetric model can be 
constructed, but eventually we will see that 
many more interesting possibilities exist for supersymmetric models.

\subsection{The Standard Model in the bulk}

We have seen that the massless modes of a 5D scalar or fermion 
field in the AdS slice can be localized at either of the two boundaries,
depending on the value of the dimensionless mass parameter $ c$.
This continues to be true for massless fermions, even without supersymmetry,  
due to the fact that chiral symmetry protects the massless mode. 
However, without supersymmetry, the massless mode of the bulk scalar 
field will receive radiative corrections of order the Planck scale. Thus,
it would appear that the hierarchy problem is again reintroduced, as in flat
space. This dire conclusion is not in fact realised, because the nonzero 
Kaluza-Klein scalar modes are localized near the TeV-brane and 
consequently have TeV scale masses. Thus, even though the scalar zero mode 
becomes heavy, the mass of the lightest nonzero Kaluza-Klein mode
is still of order the TeV scale, and therefore the Higgs can be associated
with this lightest Kaluza-Klein state. This is effectively like having the 
Higgs field on the TeV-brane.

\begin{figure}[ht]
\centerline{ { 
\begin{picture}(0,0)%
\includegraphics{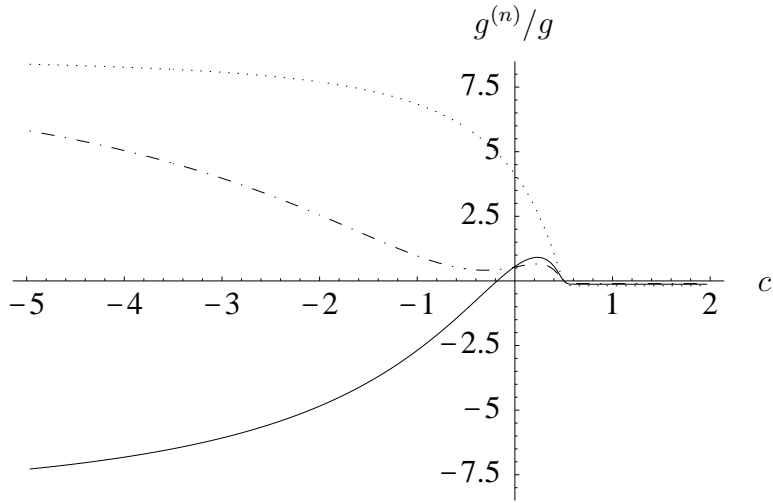}%
\end{picture}%
\setlength{\unitlength}{3947sp}%
\begin{picture}(4950,3221)(3964,-4931)
\put(7201,-2011){\makebox(0,0)[b]{$g^{(n)}/g$}}
\put(8776,-3586){\makebox(0,0)[b]{$c$}}
\end{picture}
}}
\caption{\it The ratio of the gauge couplings, $g^{(n)}/g$, for $n=1$ (dotted
line), $n=2$ (solid line) and $n=3$ (dashed-dotted line),
as a function of the dimensionless fermion mass parameter $c$.}
\label{fig:gc}
\end{figure}

If the fermion fields are located in the bulk, then the 
gauge bosons must necessarily reside in the bulk. 
In this case, the Kaluza-Klein excitations of the gauge bosons 
can induce
four-fermion interactions that are severely constrained by the experimental data
\cite{ab,em,dpq}.
To study these constraints in our case,
let us consider
the five-dimensional gauge coupling between the bulk gauge boson and fermion.
It is given by
\begin{equation}
     \int d^4 x \int dy \sqrt{-g}\, g_5 \bar{\Psi}(x,y)i \gamma^\mu A_\mu(x,y)
      \Psi(x,y)~,
\end{equation}
where $g_5$ is the five-dimensional gauge coupling.
Using the 
expression for the zero-mode fermion (\ref{eq:masslessH}), the gauge coupling
of a gauge boson Kaluza-Klein mode $n$ to the zero-mode fermions is 
\begin{equation}
   g^{(n)} = g\left(\frac{1-2c}{e^{(1-2c)\pi kR}-1}\right)\frac{k}{N_n}
    \int_0^{\pi R} dy\, e^{(1-2c)\sigma} \left[ J_1(\frac{m_n}{k} e^\sigma)
    +b_1(m_n)Y_1(\frac{m_n}{k} e^\sigma)\right]~,
\end{equation}
where $g=g_5/\sqrt{2\pi R}$ is the four-dimensional gauge coupling, 
and $c$ is 
defined in Eq.~(\ref{masses}). 
In Fig.~\ref{fig:gc},
we show the ratio $g^{(n)}/g$, for $n=1,2,3$, as a function of $c$. 
When $c$ is large and negative, the fermion is localized near the TeV-brane
and the ratio $g^{(1)}/g$ approaches the asymptotic limit 
$g^{(1)}/g\simeq\sqrt{2\pi kR}\simeq 8.4$, which corresponds to a TeV-brane 
fermion~\cite{dhr,gauge}.
In this region the first excited Kaluza-Klein mode of the gauge boson couples 
strongly to the fermion zero-mode, leading to a restrictive 
lower bound on the first excited Kaluza-Klein
mass scale $m_1\gappeq 20$ TeV \cite{dhr}. 
At $c=0$, we recover the massless fermion case considered in 
Ref.~\cite{yama}. As the value $c$ of the bulk fermion mass increases, 
we find that the bound considerably weakens, because the fermion zero mode 
couples less strongly. At the conformal limit $c=1/2$, 
the coupling completely vanishes due to the fact that the five-momentum is
conserved in this limit. For $c> 1/2$, the coupling quickly becomes universal
for all fermion masses. This is because the fermions are now localized near
the Planck-brane, where the wavefunction of the Kaluza-Klein gauge bosons is 
constant.  
In this region the asymptotic limit $g^{(1)}/g \simeq -0.2$, agrees with the 
case of a Planck-brane fermion~\cite{gauge}. If all fermions have equal
five-dimensional masses, 
the bound from electroweak precision experiments on 
the first excited Kaluza-Klein mass  is
\begin{equation}
\label{bound} 
       m_1 \gtrsim 2.1 \left(\frac{g^{(1)}}{g}\right)~{\rm TeV}~,
\end{equation}
where $m_1\simeq 2.5 k e^{-\pi kR}$.
Thus we see that in the region where $c\gtrsim 1/2$, the lower bound is 
fairly innocuous.
The coupling of the higher-order Kaluza-Klein modes ($n>1$)
is always smaller than the coupling of the
first excited state, and thus gives negligible contributions to
any four-fermion operator (unless the fermion is on the TeV-brane).
As can be seen in Fig.~\ref{fig:gc},
the ratio of the couplings, $g^{(2)}/g$, and $g^{(3)}/g$,
continue to remain $c$-independent for $c\gtrsim 1/2$.

\subsection{Fermion mass hierarchy}

In the original scenario~\cite{rs}, the warp factor $e^{-\sigma}$, was 
used to solve
the gauge hierarchy problem by generating the TeV scale from the Planck scale. 
The fermion mass hierarchy problem was not addressed there 
because the fermions were confined to the brane
and the Yukawa interactions were conformally invariant. 
However, if we allow the fermions to 
live in the bulk, a similar use of the warp factor can be used to explain the 
fermion mass hierarchies. For neutrino masses this idea has been considered
in Ref.~\cite{gn}.  Here we will analyze it for  
the quarks and leptons.
Our setup is as follows.
For each fermion flavour $i$, 
we have two five-dimensional Dirac fermions $\Psi_{iL}(x,y)$, 
and $\Psi_{iR}(x,y)$, while for simplicity the Higgs $H(x)$ will be 
localized on the TeV-brane.
Thus, the five-dimensional action will have the Yukawa coupling term
\begin{eqnarray}
   &&\int d^4x \int dy\, \sqrt{-g}\,\,\lambda^{(5)}_{ij} H(x) 
   \Big( \bar\Psi_{iL}(x,y)\Psi_{jR}(x,y) + h.c.\Big)
\delta(y-\pi R) \nonumber \\
     &&\equiv \int d^4x \,\,\lambda_{ij} H(x) \Big( \bar \Psi^{(0)}_{iL}(x) 
     \Psi^{(0)}_{jR}(x) + h.c. + \dots\Big)~,
\end{eqnarray}
where $\lambda^{(5)}_{ij}$ are the five-dimensional Yukawa couplings and 
$\lambda_{ij}$ define the effective four-dimensional Yukawa couplings of
the zero modes, $\Psi^{(0)}_{iL}$ and $\Psi^{(0)}_{jR}$.
If each fermion field has a bulk mass term parametrized by $c_{iL}$
and $c_{jR}$, then
the fermion zero modes will develop an exponential profile which gives rise
to the following four-dimensional Yukawa couplings
\begin{equation}
\label{yc}
    \lambda_{ij}= \frac{\lambda_{ij}^{(5)}k}{N_{iL} N_{jR}} 
    e^{(1-c_{iL}-c_{jR})\pi kR}~,
\end{equation}
where 
\begin{equation}
\label{norm}
    \frac{1}{N_{iL}^2} \equiv \frac{1/2-c_{iL}}{e^{(1-2c_{iL})\pi kR}-1}~,
\end{equation}
and similarly for $N_{jR}$.
Note that in deriving (\ref{yc}), the Higgs must be rescaled by an amount 
$H(x)\rightarrow e^{\pi kR} H(x)$, in order to obtain a canonically normalized 
kinetic term. 
We see then from Eq.~(\ref{yc}), that exponentially-small Yukawa couplings can
be generated for values of $c_{iL}$ and $c_{jR}$ slightly larger than
$1/2$.

\begin{figure}[ht]
\centerline{ {
\begin{picture}(0,0)%
\includegraphics{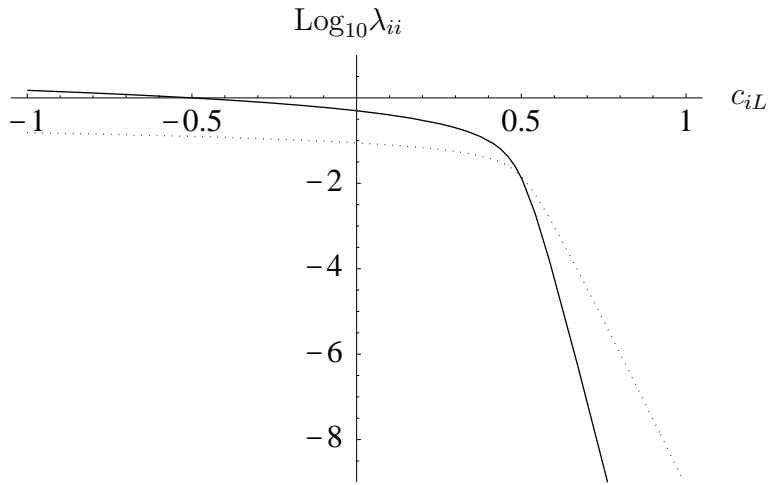}%
\end{picture}%
\setlength{\unitlength}{3947sp}%
\begin{picture}(4824,3057)(1789,-4306)
\put(3976,-1486){\makebox(0,0)[b]{${\rm Log}_{10}\lambda_{ii}$}}
\put(6376,-1936){\makebox(0,0)[lb]{$c_{iL}$}}
\end{picture}
}}
\caption{\it The effective four-dimensional Yukawa coupling $\lambda_{ii}$
as a function of $c_{iL}$,
for the case $c_{iL}=c_{iR}$ (solid line) and $c_{iR}=1/2$ (dotted line)
.}
\label{fig:yc}
\end{figure}

The precise flavour structure of the Yukawa coupling matrix,
$\lambda_{ij}$, will depend on the values of $c_{iL}$ and $c_{jR}$.
We can consider
two simple limits, which in some sense represents the two possible extremes
for the flavour structure. Suppose first that we have a left-right symmetric
theory such that $c_{iL}=c_{iR}$. 
In Fig.~\ref{fig:yc} 
we have plotted the diagonal element of the Yukawa
matrix, Eq.~(\ref{yc}), in this left-right symmetric limit.
We have assumed, for simplicity,  that $\lambda_{ij}^{(5)}k \sim 1$. 
For $c_{iL}>1/2$, we clearly see in Fig.~\ref{fig:yc} 
the exponential damping of Eq.~(\ref{yc}), where 
the diagonal elements of Eq.~(\ref{yc}) simplifies to (for $kR\gg 1$)
\begin{equation}
\label{y1}
       \lambda_{ii} \simeq \lambda_{ii}^{(5)}k\,
       (c_{iL}-1/2)\, e^{-2(c_{iL}-1/2)\pi kR}~.
\end{equation}
An electron Yukawa coupling $\lambda_e \sim 10^{-6}$ can be generated for 
$c_{eL} \simeq 0.64$ (assuming $kR \sim 12.46$). 
On the other hand, a top Yukawa coupling $\lambda_t\sim 1$ can be
obtained for $c_{tL}=-1/2$, 
since the exponential factor in (\ref{yc}), disappears 
for $c_{iL}<1/2$. 

Alternatively, we can suppose that the right-handed fermions 
all have $c_{jR}=1/2$. Now, the effective Yukawa couplings, 
in the limit that $kR\gg 1$ and $c_{iL}>1/2$ 
become
\begin{equation}
\label{yclimit}
    \lambda_{ii} \simeq 
\frac{\lambda_{ii}^{(5)} k}{\sqrt{2\pi kR}} \sqrt{c_{iL}-1/2}\,
    e^{-(c_{iL}-1/2)\pi kR}~.
\end{equation}
In this case the exponential factor is more dominant compared to
the left-right symmetric assumption. This can clearly be seen in 
Fig.~\ref{fig:yc}.

It is easy to understand  the behaviour of the curves in Fig.~\ref{fig:yc}.
For $c_{iL}> 1/2$,
the Yukawa couplings quickly become exponentially-small 
since fermions are localized near the Planck-brane and
have an exponentially small overlap with the Higgs, which is located 
on the TeV-brane. 
In this way we see that the mass of the lightest fermion states of the 
SM can be naturally generated for values of $c_{iL}>1/2$. 
The heavier fermions $(c,\tau,b)$ require a larger overlap with the Higgs
and cannot be confined near the Planck-brane.
In this case the values of the mass parameters must be close to the 
conformal limit $(c_{iL} \simeq 1/2)$, where
fermions are not localized on either of the two branes.
The top quark having a comparable mass to the VEV of the Higgs,
needs the largest overlap of all, and must be localized 
on the TeV-brane $(c_{tL} \lesssim -1/2)$. 
Again we must  stress that this scenario 
does not predict the fermion mass spectrum
(which would correspond to predicting the values of $c_{iL}$ and $c_{jR}$).
However, it does offer a possible explanation for 
the mass hierarchy between fermion families.
This idea is similar to Ref.~\cite{as}, where the 
exponential overlap between the Higgs and fermion wavefunctions also generated 
fermion mass hierarchies.

It is important to realise, however, that 
if the fermions are localized at different points (by assuming that 
they have different 5D masses $c_i$), 
then they will couple differently to the Kaluza-Klein gauge bosons. 
This can give rise to large flavour-changing neutral 
current (FCNC) processes~\cite{dpq}. 
For example, in the model of Ref.~\cite{as}, 
a lower bound on the compactification scale of $25-300$ TeV is 
obtained~\cite{dpq} (the range in  the bound depends on whether or not
the induced FCNC processes violate the CP symmetry). 
This would also seem to be a problem for our scenario.
However, in our scenario the FCNC constraints 
lead to a much smaller lower 
bound on the first excited Kaluza-Klein mass  $m_1$. For the case
represented by Eq.(\ref{y1}) we have $m_1\gtrsim 2-30$ TeV, while 
$m_1\gtrsim 0.2-2.5$ TeV in the case of Eq.(\ref{yclimit}).
The lower bounds are much reduced because, as shown in Fig.~\ref{fig:gc},  
the Kaluza-Klein gauge-bosons
couple almost universally to any fermion with $c_{iL}\gtrsim 1/2$.
This is the case for the first and second family.
The third-family
fermions do not have a universal coupling, since $c_{iL} < 1/2$, but in this 
case the FCNC constraints are much weaker~\cite{dpq}.

\subsection{Higher-dimensional operators}

When fermions are confined on the TeV-brane, as in the original 
setup~\cite{rs},
higher-dimensional operators are suppressed by the TeV scale and not
the Planck-scale~\cite{ddg}. 
In the absence of any discrete symmetries, 
this of course 
leads to proton decay problems. In addition, constraints
on $K-\bar K$ mixing
requires that the dimension-six operator, $(d\bar s)^2/M_4^2$ needs to
be suppressed by
at least an effective four-dimensional mass scale $M_4\sim 1000$ TeV.
There is also the problem of generating large
neutrino masses.

If the fermion fields are instead in the bulk, then we can ask 
whether effective
higher-dimensional operators are suppressed. 
As we saw in Section 3, each fermion zero mode has a wavefunction proportional
to $e^{(2-c_i)\sigma}$,
and depending on the values of $c_i$ can lead to an exponential suppression
of mass scales.
Let us consider first the following
generic four-fermion operators which are relevant for proton decay and
$K-\bar K$ mixing:
\begin{equation}
\label{pd}
      \int d^4 x \int dy~\sqrt{-g} \frac{1}{M_5^3} \bar\Psi_i\Psi_j
      \bar\Psi_k\Psi_l
     ~\equiv~ \int d^4 x \ \frac{1}{M_4^2} \bar\Psi_i^{(0)}\Psi_j^{(0)}
       \bar\Psi_k^{(0)}\Psi_l^{(0)}\, ,
\end{equation}
where the  effective four-dimensional mass scale $M_4$ is 
\begin{equation}
\label{m4}
      \frac{1}{M_4^2}= \frac{2k}{M_5^3}\frac{1}{N_i N_j N_k N_l}
      \frac{e^{(4-c_i-c_j-c_k-c_l)\pi kR}-1}{(4-c_i-c_j-c_k-c_l)}~,
\end{equation}
and $N_i$ are defined in Eq.~(\ref{norm}). 
For the values $1/2 \lesssim c_i \lesssim 1$, 
the four-dimensional suppression scale, $M_4$ ranges 
from the TeV scale (for $c_i \simeq 1/2$) to $M_P$ (for $c_i \simeq 1$).
Therefore, we can only obtain
the necessary suppression for proton decay if 
the mass parameters satisfy $c_i\gtrsim 1$.
This, of course, means that the light fermions are localized on the 
Planck-brane. Unfortunately, for  $c_i\gtrsim 1$
fermion masses are much too small (see Eq.~(\ref{yc})).
This is because (in non-supersymmetric models), the Higgs field
must live on the TeV-brane
and will have a very small overlap with the fermions.
Thus, it is impossible to suppress proton 
decay operators and have  fermion masses.
On the other hand, we find that 
the constraints on  $K-\bar K$ mixing, that require $M_4\gappeq 1000$ TeV,
can be  satisfied for values of $c_i$ compatible with 
generating fermion masses.

Finally, let us consider the dimension-five operator responsible for
generating Majorana neutrino masses. Again assuming that the left-handed
fields are in the bulk we obtain
\begin{equation}
      \int d^4 x \int dy~\sqrt{-g} \frac{1}{M_5^2} \Psi^T_{iL}
C_5\Psi_{iL}H(x)H(x)
      \delta(y-\pi R)
     ~\equiv~ \int d^4 x \ \frac{1}{M_4} \Psi_{iL}^{(0)\, T}C_5\Psi_{iL}^{(0)}
       H H~,
\end{equation}
where $C_5$ is the charge-conjugation operator. 
For $c_{iL} > 1/2$, $M_4$ is given by
\begin{equation}
      M_4\simeq \frac{M_5^2}{k} e^{-2(1-c_{iL})\pi kR}~.
\end{equation}
Using the relation (\ref{yclimit}), we can relate the neutrino
masses to the fermion masses:
\begin{equation}
    m_{\nu_i} \simeq  10^2\ \lambda_{ii}^2\,\,{\rm GeV}~,
\end{equation}
and leads to neutrino mass values of
$(1, 10^3, 10^7)$ eV for $(\nu_e,\nu_\mu,\nu_\tau)$. While these
neutrino masses are not ruled out by collider experiments, they can be
problematic for cosmology. In particular, if the neutrinos are stable,
then the masses do not satisfy the bound from overclosing the universe.
This constraint is easily avoided if the neutrinos decay, but this now 
requires a more detailed analysis.

In conclusion, the requirement that the Higgs field is localized 
on the TeV-brane, in order to solve the gauge-hierarchy problem, is in direct 
conflict with proton decay, although FCNC constraints can be safely 
avoided if fermions live in the bulk.
To delocalize the Higgs from the TeV-brane 
and allow new scenarios we need supersymmetry.

\subsection{Supersymmetric models}

In the case of non-supersymmetric theories we were forced to have the Higgs
field localized on the TeV-brane, with a small overlapping 
on the Planck-brane. 
However, if supersymmetry is realized, we can relax the above assumption.
The Higgs fields can now live anywhere in the bulk since
 their masslessness will be stable from radiative corrections.
We will consider two new possibilities. The Higgs can be either

a) not localized at all, 
$ c=1/2$ in Eq.~(\ref{eq:masslessH}) 
(i.e. conformally flat limit), or

a) localized on the Planck-brane, 
$ c\gg 1/2$ in Eq.~(\ref{eq:masslessH}).

\noindent
These two possibilities, (a) and (b), allow the fermions
(and their superpartners) to be placed on the Planck-brane. 
The main motivation for having the SM fermion living on 
the  Planck-brane is to avoid dangerous higher-dimensional
operators that, as discussed in the previous section,
could induce large proton decay.
By placing the fermions on the Planck-brane, any higher-dimensional
operator will  be suppressed by $M_P$  \cite{gauge}
(or the slightly smaller GUT scale of
$10^{16}$ GeV). 
This occurs if either the SM fermions are actually
confined to the Planck-brane as four-dimensional fields, or they are 
bulk fields with five-dimensional masses $ c\gg 1$ (see discussion below 
Eq.~(\ref{m4})). In the latter case,
the fermions form part of the five-dimensional hypermultiplets that, 
as shown in section~\ref{sectionsusy}, are 
localized by the AdS metric on the Planck-brane. Unfortunately with
the Higgs off the TeV-brane, the Yukawa couplings are 
no longer exponentially suppressed, and
the fermion mass hierarchy cannot be 
explained by the mechanism outlined in Section 4.2.

Finally, we will consider the SM gauge-boson as a five-dimensional bulk field.
This is a necessary consequence if some of the SM fermions (or the Higgs)
live in the bulk. There is also another motivation here.
If the   SM gauge-bosons live in the bulk, they  will behave as 
messenger fields between the two branes, communicating 
any physics on the TeV-brane to the  Planck-brane.
In particular, if supersymmetry is assumed to be broken on the 
TeV-brane, they will communicate it to the Planck-brane.
But also Planckian (stringy) physics  can be communicated to 
the Planck-brane.
Note that at the TeV-brane, Planckian physics is rescaled down to TeV
energies. 
Therefore a gauge boson can be produced at the Planck-brane
with a TeV energy and then propagate to the TeV-brane.
At the TeV-brane, it will interact with 
Planck scale excitations of the gauge boson, which now have TeV masses.
Thus, we expect that in this scenario we will be able to test GUT or theories of
quantum gravity at the TeV.

\subsubsection{Supersymmetry breaking on the TeV-brane} 

Let us  assume  that supersymmetry is broken on 
the brane at $y^\ast=\pi R$. Since all mass scales at $y^\ast=\pi R$,
are rescaled to  $M_P e^{-\pi kR}$, one can generate  small
supersymmetry-breaking soft masses. 
This presents a new alternative to models with  gaugino condensation
where the small scale of supersymmetry breaking is generated 
by dimensional transmutation.

Since we assume that the SM gauge superfields 
live in the 5D bulk, they couple to the TeV-brane  
and can feel the breaking of supersymmetry
at tree-level. In particular,
gaugino mass terms can be generated
from the TeV-brane spoiling the 
supersymmetric condition Eq.~(\ref{v:susycon}). 
This will
modify the Kaluza-Klein decomposition of the gaugino fields given
in section~\ref{sectionsusy}. 
For example, the full Kaluza-Klein mass spectrum will  
be shifted up with respect to the boson sector.
There will no longer be a massless gaugino mode in the spectrum.
Although the precise value of the Kaluza-Klein masses and eigenfunctions 
will be left for future work, we can infer here some of the  consequences.

The breaking of supersymmetry on the $y^\ast=\pi R$ brane will be
communicated to the  rest of  SM particle
superfields by the gauge superfields.
Squarks and sleptons on the Planck-brane
will receive masses at the quantum level.
Since gauge interactions are flavour-independent, the 
scalar masses will be flavour diagonal, solving the supersymmetric
flavour problem.
Notice that not only the massless mode but all the Kaluza-Klein tower
will mediate the breaking of supersymmetry
similarly as in Ref.~\cite{bulk}.
Therefore 
we expect that the sparticle spectrum will
be very similar to that in theories of supersymmetry-breaking by TeV
compactifications \cite{bulk}.
In particular we expect a mass gap between the scalar masses and the 
gaugino  masses. 
This is because the scalars will receive masses
at the one-loop level 
\begin{equation}
  m^2_i\sim \frac{\alpha_i}{4\pi} ({\rm TeV})^2~,
\end{equation}
where $\alpha_i$ is the gauge coupling of the gauge group
under which the scalar $i$ transforms. 
The scalar masses will be an order of magnitude smaller than the gaugino
masses and fulfil the relation
\begin{equation}
  \label{scalar}
  \frac{m^2_i}{m^2_j}=\frac{\alpha_i}{\alpha_j}\, .
\end{equation}
The right-handed slepton will be the lightest scalar.
However, the lightest supersymmetric particle (LSP) will be the 
gravitino. Since it couples to the TeV-brane by $1/M_P$-suppressed
interactions, its mass will be of order TeV$^2/M_P$.
This represents a very light gravitino with mass $m_{3/2}\sim 10^{-4}$ eV
and satisfies the usual constraints from cosmology and collider
experiments~\cite{grav}.

What about the Higgs?
In model (a) the Higgs will also couple to the TeV-brane 
and  Higgsino masses can be induced at tree-level.
In general, the scalar Higgs will also get masses at tree-level.
This  can be problematic since the electroweak scale will be
of order TeV \footnote{We could lower the supersymmetry breaking scale
from a TeV to  100 GeV, but in this case 
the squark and slepton masses, that arise at the loop level,
will be too small. Another alternative would be to reduce the
couplings of the  Higgs to the TeV-brane by 
increasing its 5D mass, $ c\gtrsim 1/2$, and having the Higgs
slightly localized towards the Planck-brane.}.
Therefore we must require that only the Higgsino mass is induced at the
tree-level.
The origin of such a pattern of supersymmetry-breaking 
masses must be addressed by  the fundamental  (string) theory. 
Of course, even if the  Higgs soft  masses are zero at tree-level, 
they will be induced
at the loop level.
As in the case of the squark and slepton masses, we expect
that the soft Higgs masses 
will be positive at one-loop level.
Nevertheless,  for the Higgs coupled to the top,
there are also negative two-loop effects coming from the  top/stop 
that  can dominate the one-loop  contribution and make the Higgs mass
negative \cite{bulk}.
This can lead to electroweak breaking at a scale an order of magnitude smaller
than the supersymmetry-breaking scale.
For model (b), however, the Higgsino mass is zero since 
the Higgs superfields do not couple to the TeV-brane.
This gives rise to the usual $\mu$-problem. The simple solution that we
envisage is to introduce a singlet field that couples to the Higgsinos.
This singlet can get a VEV at the electroweak scale and induce 
Higgsino masses.

Finally, we want to comment on
a different alternative to the models considered here, 
where only gravity lives in the five-dimensional bulk, while  
the complete SM, including the gauge fields, lives on the Planck-brane. 
Supersymmetry breaking arising from the brane at $y^\ast=\pi R$,
would then be communicated by gravity to the SM fields.
In this case, soft masses will be of order $(M_P e^{-\pi kR})^2/M_P$,
and implies that the radius of the extra dimension satisfies $kR\simeq 6$.
This model is very similar to 
gravity-mediated models of supersymmetry-breaking
and we do not expect any new interesting phenomenology.

\section{Conclusion}

Bulk fields living in a slice of AdS have different behaviour, 
depending on their spin.
We have studied the Kaluza-Klein mass spectrum
and wavefunctions of scalars, gauge bosons and fermions 
and analyzed the conditions for supersymmetry.
In the massless sector we find that the scalar and fermion
fields can be localized by the AdS geometry
on either brane (depending on their 5D masses),
while the gauge bosons are always nonlocalized. 
Furthermore, the gauge bosons couple to each brane
with equal strength, and therefore
provide a window for physics of the distant brane.

We have also studied the phenomenological implications of 
having the SM fields propagating  in a slice of AdS.
In non-supersymmetric theories, only the Higgs is
required to be localized on the TeV-brane.
The precise scale of the TeV-brane is constrained
by electroweak precision data, and 
strongly depends on the 5D fermion masses as shown in 
Eq.~(\ref{bound}) and Fig.~1. 
Remarkably, there is no significant bound 
if the 5D fermion masses satisfy $m_\Psi>\sigma^\prime/2$.
For these values of $m_\Psi$, the four-dimensional
massless fermions are 
slightly localized near the Planck-brane  and have
an exponentially small coupling to the Higgs.
We have shown
that this could explain the fermion mass hierarchy
by confining the light fermions to the Planck-brane,
and the heavy fermions to the TeV-brane.
Similarly, by studying 
higher-dimensional operators, we have shown that
constraints from proton decay forces
the fermions to be strongly confined 
towards the Planck-brane. 
Unfortunately, these constraints
forbid the generation of fermion masses, because now there is 
very little wavefunction overlap with the Higgs.

If  supersymmetry is present, the massless Higgs mode is 
protected from radiative corrections and the Higgs can now
live off the TeV-brane. Thus, 
higher-dimensional proton decay operators can be suppressed,
while at the same time having fermion masses.
However, the warp factor needed to explain the hierarchy
becomes impotent, since the Higgs and fermions
are located near the Planck-brane.
Assuming that the source of supersymmetry-breaking is
on the TeV-brane, the scale of supersymmetry-breaking will 
be of ${\cal O}$(TeV). This
provides a new alternative to gaugino condensation models.
If the gauge fields live in the bulk, gaugino mass terms will be
generated at tree-level.
The gauge fields will communicate the supersymmetry breaking
to the Planck-brane where the squarks and sleptons live.
Thus, the supersymmetric flavour problem is naturally solved.
This model leads to a
supersymmetric mass spectrum where the right-handed slepton is 
the lightest scalar and the LSP is a very light gravitino.
One particularly favourable model, is when the Higgs field is 
conformally flat ($c=1/2$) and nonlocalized.
In this case the Higgsino mass could also be induced at tree-level.

Finally, we reiterate that 
the whole scenario, including the details of
the source branes, must eventually 
be embedded into some underlying theory
(such as string theory). In particular,
such a fundamental theory must provide a tree-level
supersymmetry-breaking mechanism on the TeV-brane.
It is appealing that the hierarchies in the fundamental scales 
of physics can be directly related to the geometry of space-time. 
This fact alone, warrants further investigation of these scenarios.

\section*{Acknowledgements}
We wish to thank Emilian Dudas, Oriol Pujol\`as, and 
Daniel Waldram for helpful discussions. 
One of us (AP) acknowledges the ITP in Santa Barbara
for hospitality during the initial stages of this work.
The work of TG is supported by the FNRS, contract no. 21-55560.98, 
while that of AP is partially supported by the CICYT Research Project
AEN99-0766.

\end{document}